\documentclass{article}
\usepackage{graphicx}
\usepackage{geometry, amsmath, amssymb}
\usepackage{graphicx, caption, subcaption, xcolor, multicol, multirow, capt-of}
\usepackage{enumitem, lineno}
\usepackage{comment}
\usepackage{tfrupee}
\setlist{nosep}
\usepackage{abstract}

\usepackage{soul}
\usepackage[english]{babel}
\usepackage{csquotes}
\usepackage[backend=biber, style=apa]{biblatex}
\DeclareLanguageMapping{american}{american-apa}
\usepackage{authblk}
\usepackage[normalem]{ulem}

\title{Fairness, not Emotion, Drives Socioeconomic Decision Making}

\author[1,2,$\dagger$]{Rudra Mukhopadhyay}
\author[2,3,$\dagger$]{Sourin Chatterjee}
\author[2,*]{Koel Das}

\affil[1]{Department of Biological Sciences, Indian Institute of Science Education and Research Kolkata}
\affil[2]{Department of Mathematics and Statistics, Indian Institute of Science Education and Research Kolkata}
\affil[3]{Aix-Marseille Université, Inserm, Institut de Neurosciences des Systèmes, UMR 1106, Marseille, France}

\affil[$\dagger$]{Both authors contributed equally to this work.}
\affil[*]{Corresponding author: koel.das@iiserkol.ac.in}

\begin{document}
\maketitle

\begin{abstract}
Emotion and fairness play a key role in mediating socioeconomic decisions in humans; however, the underlying neurocognitive mechanism remains largely unknown. This exploratory study unraveled the interplay between agents' emotions and the fairness of their monetary proposal in rational decision-making, backed by ERP analyses {\color{black}of N170, Late Positive Potential (LPP), Feedback Related Negativity (FRN) and P3a} at a group as well as a strategic level. In a time-bound ultimatum-game paradigm, 40 participants were exposed to three distinct proposers' emotions (Happy, Neutral, Disgusted) followed by one of the three offer ranges (Low, Intermediate, High). Our findings show a robust influence of economic fairness on acceptance rates. A multilevel generalized linear model showed offer as the dominant predictor of trial-specific responses. Subsequent clustering grouped participants into five clusters, which the Drift Diffusion Model corroborates. Pertinent neural markers demonstrated the recognition of facial expressions; however, they had minimal effect during socioeconomic decision-making. Our study explores individualistic decision-making processes revealing different cognitive strategies. 
\end{abstract}



\section*{Introduction}

Human decision-making can fundamentally be imagined as a bounded optimization over perceived trade-offs \parencite{simon1972bound}. 
Within any social context, decisions are often modulated by self-interest, fairness, inherent value system, and our interpretation of others' intentions \parencite{Frith2008}. Social judgement formation is often facilitated by facial expressions which provide insights about the emotional state, intentions, and traits of others \parencite{ekman1993facial, Horstmann_2003}.

Facial signals, simultaneously, are theorised to significantly impact others' affect, cognition and behaviour \parencite{van2009emotions, vanK_2022}. Facial expressions, one of the most widely studied non-verbal cues, are strong emotional signals which can alter the perception or the emotional experience at an interpersonal level, leading to behavioural adaptations \parencite{vanK_2022, Winkielman_2022, Sagliano_2022}. A non-compliant facial display, for instance, may impact the subjective evaluation of fairness and consecutive behavioural response. 

Fairness considerations (in other words, how justly the agent themselves is being treated) are elemental in motivating cooperative or punitive judgements in social decision-making \parencite{weib2021, Sagliano_2022}.
Game theoretic paradigms, namely, the ultimatum game \parencite{guth1982experimental}, have long been utilized to study social decision making from the perspective of bargaining behaviour. In the ultimatum game (UG) framework, one person (the ``proposer") is provided with an immutable amount of money and tasked with sharing it with another participant (the ``receiver"). The recipient is forced to make a critical decision: accept or reject the suggested allocation. In case of acceptance, the money will be distributed as per the division proposed by the proposer. However, none of the parties receives any money when the receiver rejects. A similar paradigm, the dictator game (DG) \parencite{dg_forsythe_94}, is different insofar as the receiver's feedback is concerned, wherein the receiver cannot reject the dictator's offered allocation. 

Contrary to the common perception of completely rational behaviour (the ``Economic man" concept by Mill in 1874 or the rational actor model \parencite{monroe1995psychology, von2007theory}), game theory-based paradigms regularly discover that the proposer typically makes offers between 40-45\%, while offers less than about 20\% are rejected roughly 50\% of the time \parencite{knight2012fairness}. The currently accepted model suggests the notion of ``Bounded Rationality" \parencite{simon1972bound}, where instead of finding the most optimal decision, individuals choose a satisfactory option that is adequate based on other circumstantial information, including considerations regarding the context of the game, factors related to the parties involved and others. \parencite{handgraaf2003social}.  

The UG framework has witnessed numerous variants, serving as a crucial tool for investigating diverse concepts such as fairness considerations \parencite{kahneman1986fairness, von2007theory}, implicit expectations \parencite{vavra2018expectUG}, social information \parencite{bailey2013age, gender_garcia2012, gender_solnick1999, Moore2021}, facial attractiveness, expressions and interpersonal emotions \parencite{ma2015, pietroni2022interpersonal, mussel2013value, mussel2014smiling, liu2016negative, zheng2017influence, ferra2021}. Emotions, as mentioned, have emerged as a significant influencer in decision-making, prompting a profound interest in understanding how a person's emotional state can inform the decisions of receivers across various contexts \parencite{rick2008role}. 
Induced sadness has previously been shown to up-modulate rejection rates \parencite{harle2007incidental}, whereas induced happiness was found to increase acceptance tendencies \parencite{riepl2016influences}.
The perceived fairness of the offer may be hugely influenced by the emotional cues from the proposer \parencite{pietroni2022interpersonal}. If the proposer conveys a negative emotion, the receiver may think of the allocation as more unfair than the same offer made with a positive expression, possibly leading to reduced cooperation with the proposer in future interactions \parencite{murphy2022dynamic}.
In recent years, an array of studies has examined the impact of emotional displays and the reception of emotional cues from fellow players during ultimatum and dictator games. Several past studies \parencite{capra2004mood, mellers2010predicting} reported that positive emotion could increase generosity in proposers (for a UG) or dictators (for a DG), whereas other studies have shown an increase in generosity via negative emotions \parencite{tan2010happiness,forgas2013give}, and some found mixed effects \parencite{matarazzo2016effects}. The proposer usually is more generous towards compliant expressions, e.g., happy or sad, than comparatively less compliant ones, displaying anger, disgust, or neutrality \parencite{weib2021}. The effect of facial expressions on the receiver's behaviour, however, is not consistent. Mussel and colleagues observed a systematic decrease in acceptance rates as the proposer's facial expression shifted from happy to neutral and further to angry \parencite{mussel2013value, mussel2014smiling}. Liu and colleagues demonstrated that the acceptance rate decreases exclusively for the negative facial expressions, displaying sadness, disgust, and fear \parencite{liu2016negative}. However, a recent study by Ferracci and colleagues suggests a strongly fairness-driven response amongst the participants, with a small main effect of facial expression \parencite{ferra2021}. 

Reaction time analysis UG framework further illuminates the processes involved in social decision-making. Previous studies have reported an inverse correlation between reaction time and offer amount, indicating that lower offers require comparatively more processing time. In contrast, larger offers are decided upon faster due to their relative ease \parencite{harle2012social,ma2015}. Knight and colleagues showed this offer-reaction time relation can be quadratic suggesting the dilemma of mid-range offers \parencite{knight2012fairness}. 
The dilemma in decision-making introduced via a third-party candidate's attractiveness resulted in the fastest response to the condition where the subject and the candidate both received fair offers and an effect of the candidate's attractiveness in the same condition \parencite{ma2015}. 
Additionally, it has been shown that higher reaction time is correlated with a higher rejection rate, and participants spend less time accepting a fair offer than rejecting unfair offers \parencite{mussel2013value, rego2016adult}. However, no significant modulation of reaction time was observed for different facial expressions of the proposer \parencite{liu2016negative}.

The neural substrates underlying participants' response to emotional expressions of proposers and offer-triggered neural markers have extensively been studied in the context of ultimatum game \parencite{mussel2014smiling, polezzi2008mentalizing, weiss2020smiling}. Emotional correlates are analysed using P100, N170 and late positive potential. The first two markers are particularly sensitive to human face-like stimuli \parencite{mangun1995neural, bentin1996electrophysiological, rossion2000n170} and are assumed to be modulated by perceived expressions during passive viewing tasks \parencite{schindler2020attention, santos2008differential, hagemann2016too}. The LPP component indicates prolonged attention allocation on emotionally relevant stimuli \parencite{yen2010emotional, schindler2020attention}. Offer-related neural mechanisms are studied in relation to feedback-related negativity (FRN) and P3. The former is believed to be associated with punitive intentions in response to unfavourable feedback, e.g., unfair offers or negative facial expressions \parencite{polezzi2008mentalizing, mussel2014smiling}. The P3 ERP component is typically divided into two subcomponents: P3a, a fronto-central positivity linked to attentional orienting and working memory updating, and P3b, a parietal positivity associated with stimulus evaluation and memory processing \parencite{polich2007updating}. While P3a has traditionally been associated with novelty detection, salience-driven attention allocation, or response inhibition \parencite{polich2007updating}, recent work shows its involvement in decision-making contexts, including economic games. For instance, Peterburs et al. demonstrated a frontally distributed P3a response to unfair offers in the Ultimatum Game \parencite{P3_Peterburs2017}. A higher P3a component has been observed when participants switch between high- and low-risk options in a monetary gambling task \parencite{zhang2013electrophysiological}. Higher P3b amplitude indicates greater confidence in decisions and more effort devoted to a task \parencite{P3_Peterburs2017, P3_Mansor2021}. 


Prior research indicates conflicting evidence regarding the impact of proposer's facial expressions and offer fairness on acceptance rate of receivers. Additionally, there are limited studies exploring reaction time. In the current study, we endeavour to extend the existing literature on the impact of facial expression-led interpersonal emotions and the fairness of the offer on acceptance behaviour and reaction time. 
More importantly, there is an overall lack of literature which examines the individual-level variations in strategy while playing an ultimatum game. Our study aims to explore different approaches of game-play across participants, identify major strategies employed via cognitive modelling and suitable clustering techniques, explain such strategy-specific decision-making via the classical drift-diffusion modelling (DDM) \parencite{ratcliff2008ddm} and validate the findings by neurophysiological insights \parencite{Sanfey2007_neural}. Despite prior discussion on the Drift Diffusion Model's (DDM) relevance to the Ultimatum Game (UG) \parencite{wei2022dual}, this specific area has received limited research attention \parencite{Numano2025ddm}.
The decision-making in a UG can be modelled as an accumulation of the perceived priority amongst the self-interest motive that constantly aims at maximizing the payoff and the emotive aspect that considers fairness criteria, interpersonal emotions, internal expectations and other factors \parencite{Sanfey2008dual, wei2022dual}. This theoretical framework is consistent with the premise on which accumulation models like DDM are constructed. 
In the current context, we aim to understand how the offer magnitude and emotion impact the DDM parameters, including drift rate (the rate of information accumulation) and starting position (initial bias towards either of the choices). Since emotional expression precedes the offer presentation in the current study, we hypothesized that the effect of emotion would manifest in the starting position. However, participants took the decision while perceiving the offer, thus, we expected the offer magnitude to influence  the speed of reaching a decision, namely, the drift rate. A visual representation of the hypotheses is depicted in Supplementary material Figure 3.

The primary objective of this exploratory study is to investigate the interplay between fairness considerations and perception of emotional expressions in shaping rational decision-making within the ultimatum game paradigm. 
Secondly, we aim to explore reaction time as a crucial behavioural variable, examining its association with fairness perceptions and offer magnitudes.
Thirdly, we aim to construct a mixed-effect model explaining the participants' responses based on the offer magnitude and the proposer's expression. 
Fourthly, a particular focus is placed on uncovering individual-level variations in strategizing and characterizing these strategies through drift-diffusion modelling (DDM) to explain cognitive mechanisms underlying decision-making. Lastly,  we attempt to integrate neural markers such as P100, N170, LPP, FRN, and P3a to probe the neurophysiological underpinnings of the behavioural responses and individual-level variations therein.




\color{black}

\section*{Materials and Methods}
\subsection*{Participants}
Fifty participants ($N_0 = 50$, all heterosexual male), within the age range of 18 - 20, with normal or corrected eyesight and no previous psychological history, were recruited from the first-year Bachelor's cohort of the institute conducting the experiment. Participants were barred from participating in the experiment if they had any earlier exposure to game theory and different types of trust games. After obtaining instructions and reading the procedural details, the participants provided written consent for their participation. The institute's Ethics Committee approved all procedures and protocols.
Ten participants were discarded due to the poor EEG recording. The final sample size of the analysis was forty male adults ($N = 40$).
The sample size is justified for the sample size calculation using G*Power 3.1, suggesting a required size of 39 (effect size partial $\eta^2 = 0.08$, $\alpha = 0.05$, $3 \times 3 \times 4$ repeated-measure design) \parencite{gpower}. Two works, which essentially motivated this study, showed the effect size of facial expressions to be widely different: $\eta_p^2 = 0.04$ \parencite{ferra2021} and $\eta_p^2 = 0.1$ \parencite{mussel2013value}). For the current study, we adopted an effect size ($\eta_p^2 = 0.08$) intermediate between these values, corresponding to a level slightly above a medium effect (0.06).

\subsection*{Experimental Paradigm}
After the rules of the ultimatum game were explained, the participants were informed that they would be partaking in an online game-based experiment. They would always play as the \textit{receivers}, whereas players from an anonymous partner institute in India would assume the roles of \textit{proposers}, even though the offers were actually computer-generated. Each participant played the game for three hundred and sixty (360) trials using a repeated-stimulus design.
The participants were also told that photos of both \textit{receivers} and the \textit{proposers} would be taken using web cameras, which would be exchanged. Consequently, in each trial, they would see a photo of their opponent on the screen, with whom they were connected for that specific trial. This pairing would renew after completing each trial. 
These photos, however, were sourced from the Indian Spontaneous Emotion Database \parencite{ised}, containing near frontal photos of fifty (50) individuals, each annotated with the emotion displayed by the individual. We selected twenty-four (24) male individuals. For each individual, we used eight photos portraying each of the following emotions: ``Happy," ``Disgusted," and ``Neutral" (or expressionless). These photos were further processed in Adobe Photoshop $^\copyright$, for centering and creating a uniform background. We avoided using photos where the head tilt and rotation angle were noticeable. 

\subsection*{Procedure}
Firstly, the participant was asked a few preliminary questions to gauge their understanding of the ultimatum game paradigm being explored. Only one participant was excluded after showing some prior knowledge regarding this.

\begin{figure}[!ht]
    \centering
    \includegraphics[width = \textwidth]{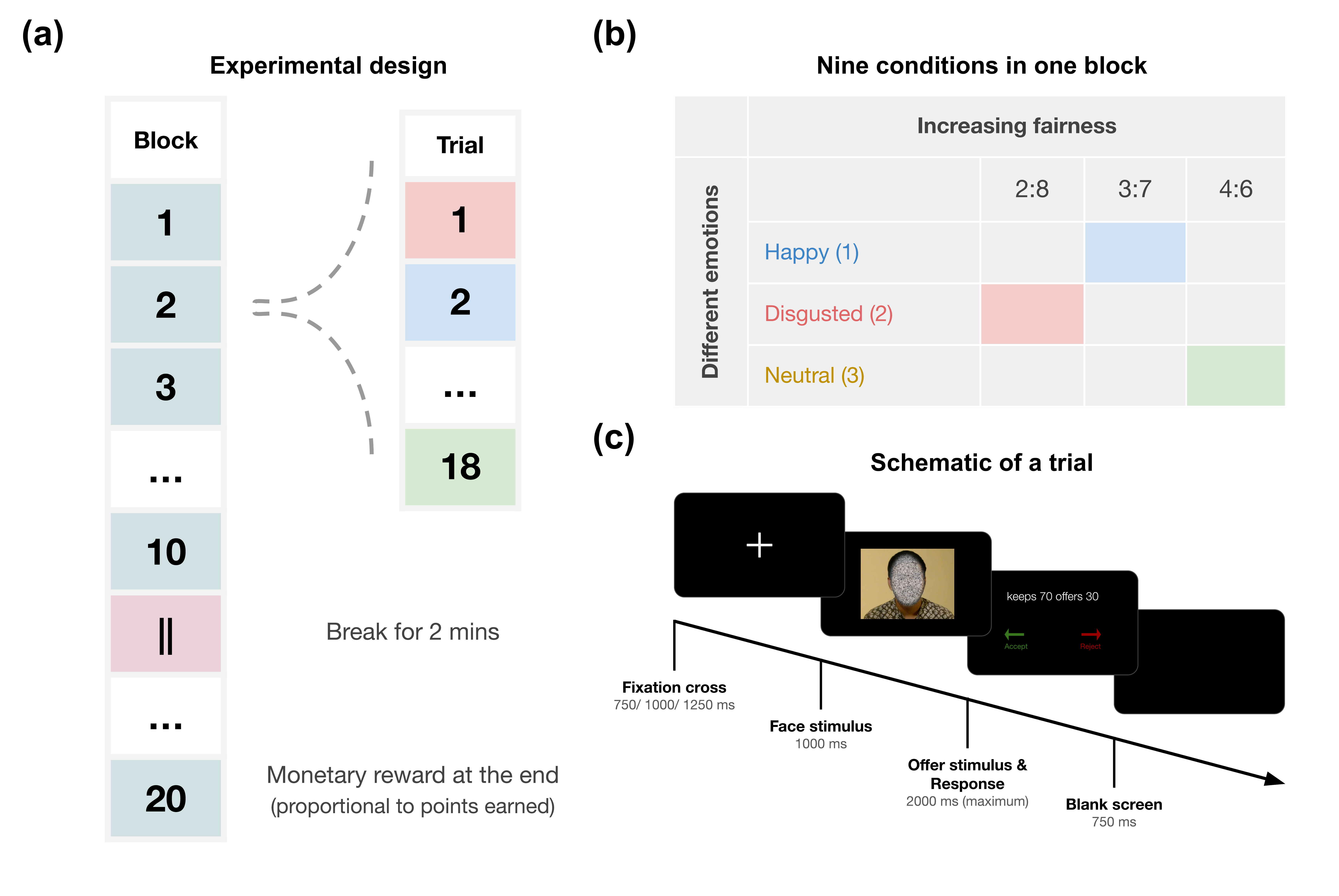}
    \caption{(a) \textbf{Experimental procedure.} The experiment was conducted in twenty (20) blocks, each with eighteen (18) trials. A mandatory two-minute break was provided after ten (10) blocks. (b) \textbf{Conditions in one block.} Nine (9) conditions, by combining three (3) kinds of emotion stimuli and three (3) offer ranges in all permutations, were repeated twice. (c) \textbf{Schematic of one EEG trial.} A fixation screen with a centre cross was displayed with a jittered time window before the appearance of face stimuli. Following proposer's face, the offer was presented in textual format. The participant would accept or reject the offer by pressing the right or left arrow key (randomized).}
\end{figure}

During the experiment, each participant played three-hundred and sixty (360) trials, which were segmented into twenty (20) blocks. A short break was provided between consecutive blocks, and a mandatory two-minute break was given after every ten blocks.
The offers (coming from the virtual proposers) were based on three different ratios: $20:80, 30:70 \text{ and } 40:60$, each favouring the proposer. The actual offer was generated by uniformly sampling three intervals: $15-24, 25-34 \text{ and } 35-44$, where the amount of money to be divided was \rupee 100 (Table 1). As mentioned earlier, the facial expression of the proposer could be one of the following: ``Happy," ``Disgusted," and ``Neutral". In each block, each of the nine (9) permutations of the three offer ratios and three emotional expressions was iterated twice (Figure 1 (a) and (b)).

\begin{table}[!ht]
    \centering
    \captionsetup{width=\linewidth}
    \begin{tabular}{c c c c c}
        \textbf{Ratio} & \textbf{Proposer receives} & \textbf{Proposer offers} & \textbf{Proposer retains} & \textbf{Code}\\
        \hline
        \hline
        2:8 & 100 & 15 - 24 & 76 - 85 & Low\\
        
        3:7 & 100 & 25 - 34 & 66 - 75 & Intermediate\\
        
        4:6 & 100 & 35 - 44 & 56 - 65 & Maximum\\
        \hline
    \end{tabular}
    \caption{Possible offers from the proposers: a range denotes that the offer is generated by uniform random sampling over that interval.}
\end{table}

\noindent
Each trial began with a fixation cross at the center of the screen for a jittered time span (750, 1000 or 1500 ms). Following this, the face of the proposer was presented for 1000 ms. After that, the proposer's offer was displayed in textual format. In this screen, the participants had 2000 ms to respond by pressing the left or right arrow key, thus ending that particular trial. If no response was recorded within the stipulated time, the trial was labelled as a ``miss" and was later excluded from further analysis. Trials thus rejected comprise approximately $0.7\%$ of the entire data.\\ 
The response-key correspondence was randomized across trials to avoid association bias. The trial ended in a blank screen that remained for 750 ms. After each block, participants received feedback on their net earnings for that block.
Following to the experiment, participants filled out an Attractiveness Rating Questionnaire, rating the expressionless photos of the proposers they played against on a scale of 1 to 10. In the end, participants received an endowment of \rupee 100  and a bonus amount proportional to their total score in the game. 

\subsection*{Behavioural Data Collection and Analysis Techniques}
Two behavioural parameters were recorded during the experiment: acceptance value -- whether the trial results in acceptance or not $(r_{i,j} = \text{1, (if accepted), 0 (otherwise) for the } i^{\text{th}} \text{ participant and } j^{\text{th}} \text{trial})$ and the reaction time $(rt_{i,j} \text{ for the } i^{\text{th}} \text{ participant and } j^{\text{th}} \text{trial})$.
\\
The acceptance value was used to estimate the acceptance rate $\displaystyle \rho_{i, \mathcal{N}} := \big(\sum_{j \in \mathcal{N}} r_{i,j}\big) / |\mathcal{N}|$, where $i = \text{ participant index}, \text{ } j = \text{ trial index}, \text{ } \mathcal{N} = \text{trials considered to calculate the acceptance rate}$.  Also, $|\mathcal{N}|$ denotes the number of trials in set $\mathcal{N}$. This set ($\mathcal{N}$) was constructed based on one of the nine (9) conditions and a time consideration (detailed discussion in the following section). 
\\
Before conducting any test or analysis, the reaction time was first filtered with a lower bound at 250 ms (accepted to be a minimum time required to generate any voluntary movement in human \parencite{Myers_2022}) and an upper bound at the upper 99-th percentile of the observed reaction latency. These trimming removed $<1\%$ of the total trials across all participants.
To analyse the reaction time, we calculated the mean of the log-transformed reaction time across all the trials in the set $\mathcal{N}$ for each participant, denoted by $\displaystyle \bar{rt}_{i,\mathcal{N}} := \big(\sum_{j \in \mathcal{N}} \log rt_{i,j}\big) / |\mathcal{N}|$ (following the notation defined earlier).  All the statistical tests performed on these two parameters assumed a repeated-measure design.

\subsubsection*{Comparing Acceptance Rate and Reaction Time Across Conditions}
In this design, all the participants were randomly but repetitively exposed to each of the nine conditions (forty times during the experiment, thus making up a total of 360 trials for each participant). A repeated-measure ANOVA was used to test whether the acceptance rate ($\rho$) varied across nine emotion-offer combinations, assuming a three-way design with three emotions shown by the virtual proposers, three ranges of offers and block-quarter (mutually exclusive cluster of four consecutive blocks, e.g., $\{1,2,3,4\}, \{5,6,7,8\},$ etc.) as a proxy for ``time". 
Similarly, a three-way repeated-measures ANOVA was performed to compare the mean reaction time ($\bar{rt}$) across the same nine conditions. 

\subsubsection*{Cognitive Modelling}
We constructed a multilevel mixed-effect linear model to predict single-trial acceptance behaviour from the emotion displayed, offer magnitude and reaction time. However, reaction time was found to be a function of emotion and offer, thus confounding the response. Thus, we considered a ``residual reaction time", devoid of the effect of emotion and offer, to model the response. The reaction time was filtered and log-transformed (as discussed previously). \\
We tested nineteen different modelling frameworks and selected the best schema based on their corrected AIC ($\text{AIC}_c$) \parencite{hurvich1989}. Details about all the schemas can be found in the Supplementary Material. Only the best-performing one is presented below using the Wilkinson notation \parencite{Wilkinson1973}.
\[
\begin{aligned}
&\textbf{Stage 1 (Reaction Time model):} \\
&\hat{rt}_{ibej} = 1 + \mathcal{O}_{ibej} 
+ (1 + \mathcal{O}_{ibej} | i) 
+ (1 + \mathcal{O}_{ibej} | i \cdot b)
+ (1 | i \cdot e)
+ (ar1_{j-1} | i)\\
\\
&\textbf{Stage 2 (Calculation of residual RT)}\\
&rt'_{ibej} = rt_{ibej} - \hat{rt}_{ibej} \\
\\
&\textbf{Stage 3 (Response model):} \\
&\mathcal{L}(\hat{r}_{ibej}) = 1 + \mathcal{O}_{ibej} + rt'_{ibej}
+ (1 + \mathcal{O}_{ibej} + rt'_{ibej} | i)
+ (1 + \mathcal{O}_{ibej} + rt'_{ibej} | i \cdot b)
+ (1 | i \cdot e)
+ (ar1_{j-1} | i)
\end{aligned}
\]
In this three-stage hierarchical model, the first equation models the log-transformed response time $rt_{ibej}$ for subject $ i $, block $ b $, emotion condition $ e $, and trial $ j $ as a linear function of normalized offer $\mathcal{O} = (o - \mu_o)/ \sigma_o, \text{ } o \in \{15, 16, \dots 44\}$, treated as a continuous variable. We check the random effect on the intercept at three levels, namely, the participant $(i)$, block-quarter index for each participant ($b$) and emotion for each participant ($e$). The random effect on the slope is checked only for the first two grouping factors. Additionally, a first order autocorrelation is included which would vary across participants.
\\
Similar is the case for the last level, in other words, the `Response model'. Here, the logit-transformed acceptance value (denoted by $\mathcal{L}(\hat{r}_{ibej})$) is modelled as a mixed linear model on normalized offer ($\mathcal{O}$) and residual reaction time ($rt'$), keeping rest of the formalisation (in other words, the random effect terms and the autocorrelation term) unaltered.\\
Following the model construction and validation (using 5-fold cross-validation), we proposed an agglomeration hierarchical approach to cluster the participants based on their strategies, represented by the coefficients corresponding to emotion and offer variables in the response model. All computations were performed using R and MATLAB$^\copyright$.

\subsection*{DDM-based Analysis}
A simple DDM-based analysis was conducted on the entire dataset using cluster identity (as retrieved by the agglomerative clustering, mentioned in the earlier subsection) and participant identity as two grouping factors. The parameters include drift rate ($v$), starting bias ($z$), decision threshold ($a$) and non-decision time ($t_{non}$). The drift rate was modelled as a Bayesian regression of the emotion index and normalized offered magnitude. In contrast, the starting point was considered to be dependent on the emotion index, as emotion is the only factor to set any predisposition prior to seeing the offer. The model schematic is provided below.

\[
\begin{aligned}
& v_{i,k} \sim \mathcal{N}(v_{0i} + v_{ek} + v_{ok} \mathcal{O}, s^v)\\
& \mathcal{L}(z_{i,k}) \sim \mathcal{N}(z_{0i} + z_{ek}, s^z)\\
& \log a_{i, k} \sim \mathcal{N}(a_{0i} + a_{ek} + a_{ok} \mathcal{O}, s^a)\\
& \log t_{non, i, k} \sim \mathcal{N}(t_{non,0,i}, s^{t_{non}})\\
& rt_{i, k} \sim \text{FPTD}(v_i, z_i, a_i, t_{non,i})\\
\end{aligned}
\]

Notation: $i$ denotes the participant index whereas $k$ is the cluster index. $v, z, a \text{ and } t_{non}$ denote the drift rate, starting point, decision boundary and the non-decision time, respectively. $v_{0i}$ is the intercept with a random effect of the inter-participant variability ($i$), $v_{ek}$ refers to the effect of emotion ($e$) with cluster identity ($k$) being the random factor. On the contrary, $v_{ok}$ is the effect of normalized offer $\mathcal{O}$ with cluster identity being the random factor. All other coefficients ($z_{0i}, z_{ek}, a_{0i}, a_{ek}, a_{ok} \text{ and } t_{non, 0, i}$) can be interpreted similarly.\\
The drift rate regression has an identity link function. The starting bias regression is logistic linked (denoted by $\mathcal{L}$), whereas threshold and non-decision time have natural log ($\log$) as the link function. Also, $\mathcal{N}(., s^{(.)})$ is a normal distribution with variance $s^{(.)}$ and $FPTD$ stands for the first passage time distribution \parencite{Smith2023}.
Note that the fitting and following analyses were performed using the Python package HSSM (Hierarchical Sequential Sampling Modeling) \parencite{hssm}.
A working hypothesis and the direction of the effect are visually depicted in Supplementary material Figure 3.

\subsection*{Neural Data Acquisition and Preprocseeing}
EEG activity was recorded using 64-channel active shielded electrodes mounted in an EEG cap following the international 10/20 system. We have used two linked Nexus32 bioamplifiers at a sampling rate of 512 Hz to record the EEG signals. Trials were epoched using the available trigger information and preprocessed using the EEGLAB toolbox \parencite{delorme2004eeglab}. The epoched data are  band-pass filtered in the 0.01 - 40 Hz range, followed by baseline correction. Bad channels were subsequently deleted using EEGLAB's Channel rejection tool (Spectrum based method with Z-score threshold=4). Rejected channels were interpolated and noisy trials were identified and statistically improbable trials were removed with probability based measure using EEGLAB's trial rejection menu (single channel Z-score threshold=8, all channel Z-score threshold=4). Finally, EEG signals were average referenced and Independent Component Analysis (ICA) was performed using EEGLAB. Ocular, muscular,
or any other artefacts were rejected based on the MARA extension operating on ICA labels \parencite{winkler2011automatic}.


\subsection*{Neural Data Analysis Techniques}
EEG-based analysis was employed primarily to explore two questions. The first was whether the participants registered the proposer's faces when the face stimuli appeared on the screen. If they did, was there differential attention corresponding to different facial expressions? The second one was whether the participants were cognizant of the offer's ``fairness" \parencite{polezzi2008mentalizing} at the neural level. Accordingly, the EEG signals are epoched twice: once based on when the face stimuli appeared (Figure 6) and secondly when the offer was displayed (Figure 7). Peaks or time-averages of the data were calculated from the average ERP over trials for an individual subject and a specific condition. \\
We analysed the P1 component at the occipital electrodes (O1, O2, Oz), the N170 component at the parieto-occipital electrodes (PO7, PO8, P7, P8), and the LPP component at parietal and parieto-occipital electrodes (Pz, P1, P2, P3, P4, POz, PO3, PO4) \parencite{schindler2020selective} for the face triggered ERP.
A standard peak detection routine was used to detect the P1 peak in 90 ms to 160 ms time interval \parencite{herrmann2005early} and the N170 peak in 150 ms to 220 ms time interval \parencite{johnston2015n170}. Early, middle, and late portions of the LPP component were calculated as time averages from 400 ms to 600 ms, 600 ms to 800 ms, and 800 to 1,000 ms post-stimulus \parencite{liu2012neural}.\\
We used the standard ERP marker- Feedback-Related Negativity, to investigate the neural characterization of perceiving the offer's fairness for offer-triggered ERP. We use the average interval of 270 ms to 310 ms to find the FRN amplitudes at the F1, F2, Fz, and Fcz electrodes \parencite{gehring2012error, mussel2014smiling}. In the same electrodes, we also investigated the P3a component in 310 ms to 330 ms time intervals. In the 330 ms to 400 ms time interval we tried to detect P3b component at parietal and parieto-occipital electrodes (Pz, P1, P2, P3, P4, POz, PO3, PO4). {\color{black} Importantly, we did not include the Reward Positivity (RewP) component in the current study even though it is theoretically similar to the FRN. RewP is commonly detected around 250–350 ms and is frequently associated with processing outcomes that are better than expected \parencite{Bauer2022rewp}.  Generally speaking, the RewP and FRN are two interpretations of a common frontocentral assessment process, where the RewP represents better-than-expected results and the FRN represents worse-than-expected results \parencite{Tunison2019rewp}. However, ``better-than-expected" and "non-reward" outcomes are not meaningfully contrasted in our paradigm's feedback structure.  The interpretability of a reward-positivity perspective is limited because every offer deviates from an equitable (50–50) distribution.  Because the FRN framework better captures the paradigm's offer design, we concentrate on it here.}

A one-way repeated measure ANOVA was employed to test all the ERP components mentioned above because of the iterative design of the experiment. A post hoc test was performed if the ANOVA indicated a significant outcome.

\section*{Results}
\subsection*{Effect of Facial Expression and Offer on Acceptance Rate}
A three-way $3 \times 3 \times 4$ (emotion, offer and time expressed as the block index) repeated measure ANOVA on the acceptance rate showed a strong main effect of the offer's fairness ($F = 60.84, \eta^2_p = 0.61, p < 0.001$). A significant effect was observed for ``time", characterized by the block-quarters, which led to an increased acceptance
rate ($F = 5.31, \eta^2_p = 0.12, p = 0.002$), indicating ``learning" more rational behaviour. Though the mean 
acceptance rate increased from a disgusted proposer to a happy one, the main effect of emotion remained insignificant ($F = 1.34, \eta^2_p = 0.03, p = 0.27$), as depicted in Figure 2 (d).
\\
No interaction was found between emotion and offer magnitude ($F = 1.01, \eta^2_p = 0.02, p = 0.40$, Figure 2 (d)). Interestingly, we found a significant interaction between time and offer magnitude in shaping the response ($F = 3.45, \eta^2_p = 0.08, p = 0.003$, Figure 2 (d)). The ``learning" effect is the strongest for the lowest range of offer and almost nonexistent for the highest range (Figure 2 (d)).

\subsection*{Effect of Facial Expression and Offer on Reaction Time}
Acceptance took less time than rejection consistently across all the conditions (verified by the KS test, $p < 0.05$ in each condition, Figure 2 (c)). To test the effect of emotion, offer magnitude and time on reaction time, we conducted a $3 \times 3 \times 4$ (emotion, offer and time expressed as the block index) repeated measure ANOVA as before. Two strong main effects were due to offer ($F = 15.49, \eta^2_p = 0.28, p < 0.001$) and time ($F = 25.21, \eta^2_p = 0.39, p < 0.001$), shown in Figure 2 (e). We also found a significant main effect of emotion ($F = 5.11, \eta^2_p = 0.12, p = 0.008$).
\\
The reaction time consistently decreased with increasing offers, which could be explained by a higher proportion of acceptance for more and more fair offers and acceptance being faster than rejection. Tukey's posthoc test showed that this effect was prominent between the intermediate and maximum ranges ($\delta_{\mu} = 0.047, p < 0.001$) as well as the low and maximum ranges ($\delta_{\mu} = 0.073, p < 0.001$). As the game progressed, the reaction time decreased, furthering the ``learning" hypothesis. According to Tukey's posthoc test, the differences between consecutive quarters were always significant ($p < 0.01$) except for the third and fourth ones ($p > 0.05$, Figure 2 (e)). The interaction between offer and time was significant with a small to moderate effect size ($F = 2.781, p = 0.012, \eta^2 = 0.07$). However, that between emotion and time was non-significant ($F = 0.941, p = 0.466$).
\\
To further investigate the effect of emotion, Tukey's posthoc test revealed that the significance is primarily due to the difference in reaction time between the disgusted and the neutral expression ($\delta_{\mu} = 0.016, p = 0.003$). The differences amongst disgusted and neutral ($p = 0.308$) or happy and neutral ($p = 0.267$) turned out to be non-significant. However, no interaction between emotion and offer was observed.
\\
These observations hint at the possibility that the participants internalized the facial expressions of their opponents during the game-play, which led to an emotional load on their decision-making (in other words, heightened processing time against emotional stimuli). 

\begin{center}
  \includegraphics[width=\textwidth]{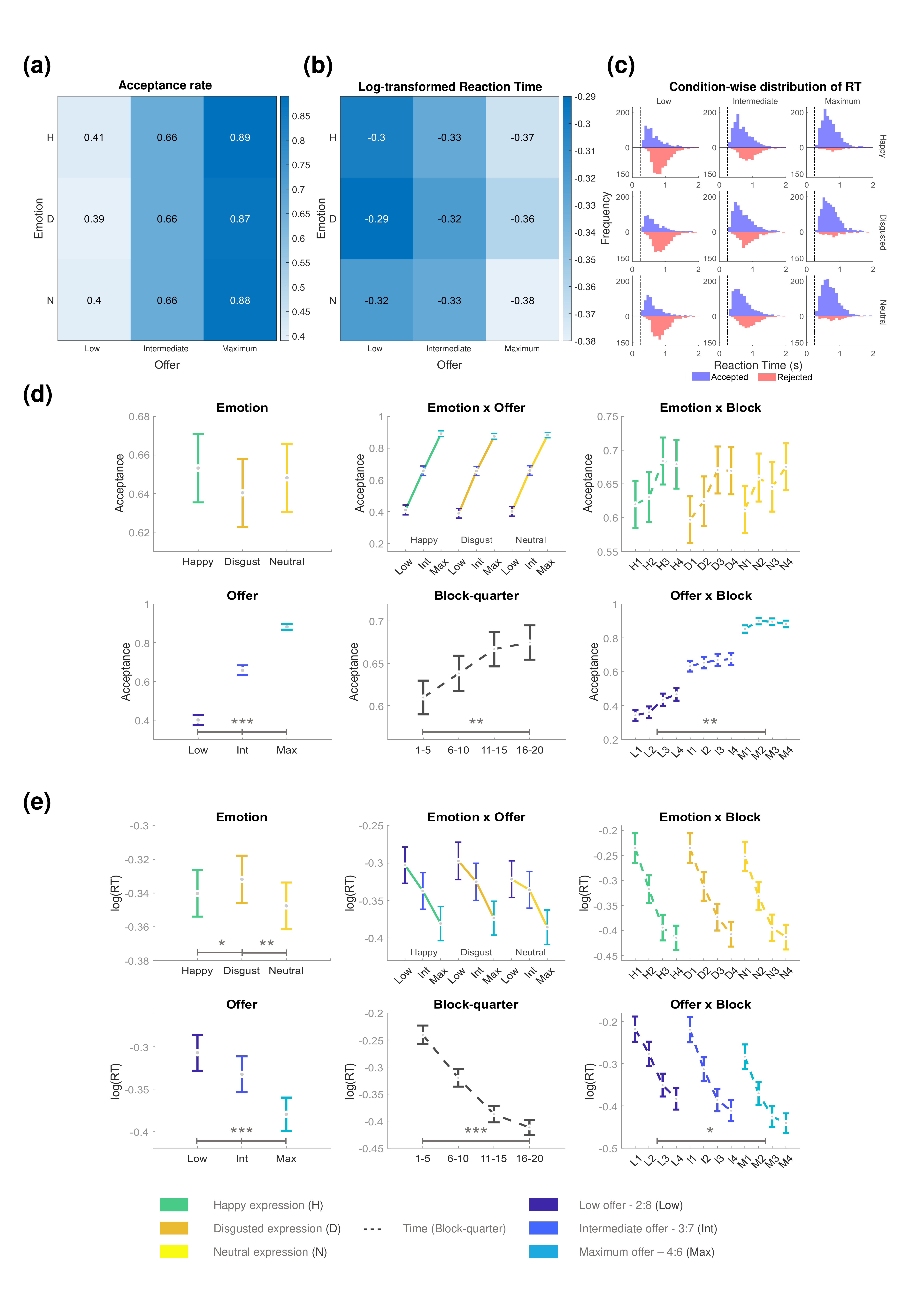}
  \medskip
  \emph{(Caption on next page.)}
\end{center}
\begin{figure}[!t]
  \caption[]{(a, b) \textbf{Heatmap} of acceptance rate and log-transformed reaction time across nine conditions. (c) \textbf{Distribution of reaction time} across nine conditions. Rejections were slower than acceptances. (d) \textbf{A repeated-measure ANOVA on acceptance rate} revealed a significant main effect of the offer's fairness and time, but not facial expression. No interaction between expression and offer was found. However, the impact of time on acceptance rate (in other words, learning) varied across offer ranges. (e) \textbf{A repeated-measure ANOVA on log-transformed RT} suggested a significant main effect of facial expression, offer's fairness and time. No interaction between offer and expression existed.\\
  (*): $p < 0.05$, (**): $p < 0.01$, (***): $p < 0.001$}
\end{figure}
\pagebreak

\subsection*{Cognitive Modelling}
\begin{figure}[!ht]
    \centering
    \includegraphics[width=\textwidth]{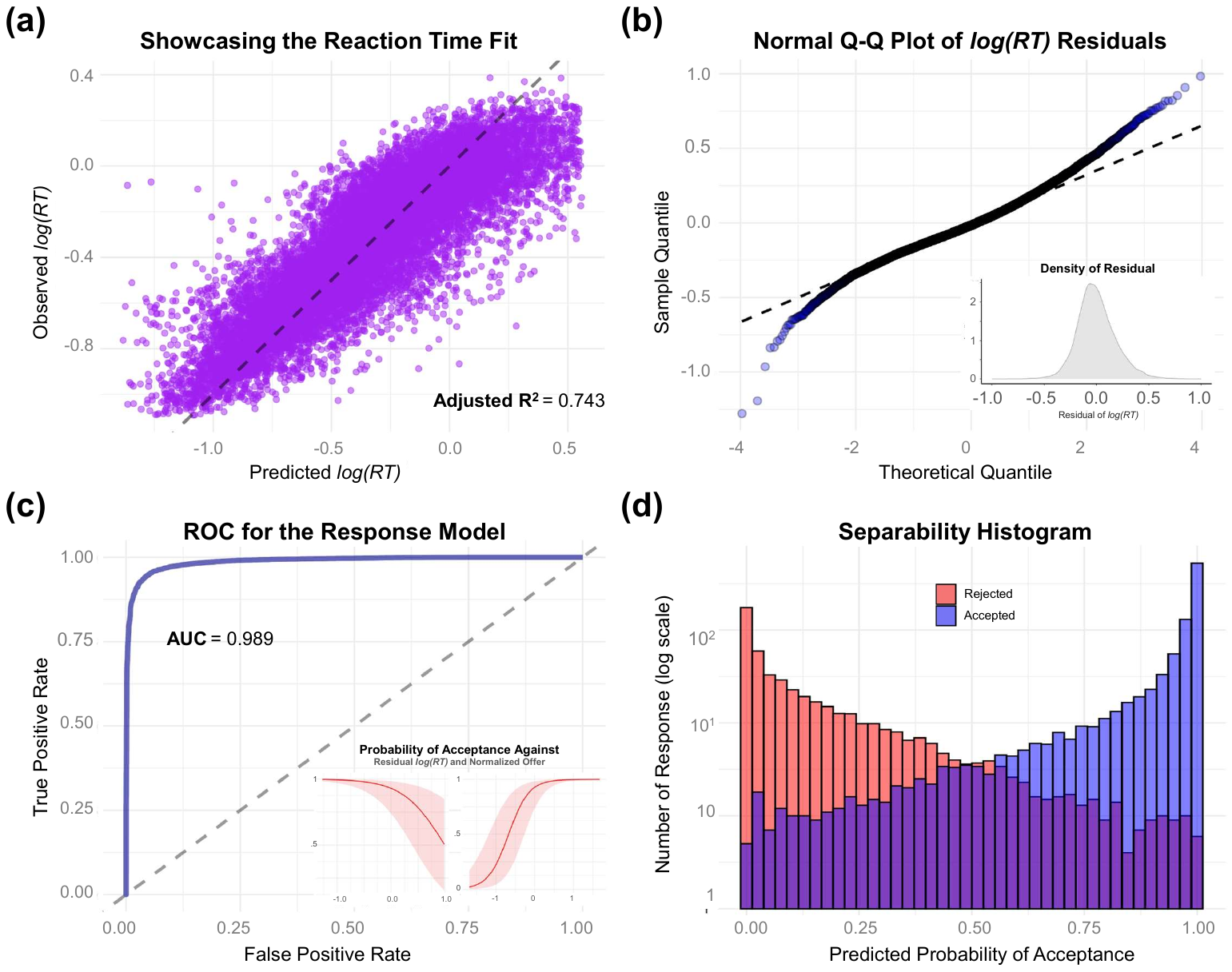}
    \caption{(a) \textbf{RT model fit.} A scatter with the observed RT (log-transformed) and predicted log-transformed RT. Except for a tendency to under-fit at higher RTs, the model performed well (Adjusted $R^2$ = 0.743 for this iteration). (b) \textbf{Residuals.} The Q-Q plot majorly follows the Q-Q line, particularly for the mid-valued reaction times. The deviations at both the ends possibly hint at the empirical distribution is more heavy-tailed than the predicted one. The density distribution of the residual is depicted in the inset. (c) \textbf{Response model performance.} The ROC curve shows the response model's predictive performance on the test set (20\% of the trials). For the current iteration, the $AUC$ value for acceptance (in other words, $\rho = 1$ is 0.989). In the inset are the marginal effect plots for the response model and the statistically significant variables, which are the residual $log(RT)$ or $rt'$ (left) and the normalized offer or $\mathcal{O}$ (right). (d) \textbf{Separability plot.} The x-axis shows the predicted probability of acceptance, and the y-axis represents the number of test trials accepted (blue) or rejected (red). Our model predicts higher probabilities for most accepted trials and vice versa, indicating effective prediction.}
\end{figure}
This section presents the results of the multilevel mixed-effects linear model described in the Materials and Methods.
\\
The outcome of modelling: the log-transformed reaction time at the first level (adjusted $R^2 \approx 0.743$, consistent over ten CV iterations, Figure 3 (a)) and the logistic response variable at the third level ($\mu(AUC_{1:=\text{accept}}) = 0.989, \sigma(AUC_1) = 0.004$ for the 5-fold CV test set, Figure 3 (c)) is presented in Table 2. Note that the estimates were derived from the model trained over the complete pooled data.
The results obtained via the modelling framework were consistent with the repeated measure ANOVA outcomes: in modelling log-transformed reaction time, emotion (a disgusted face, in particular), and offer played a significant role, while offer and residual reaction latency (not included in the ANOVA analysis), but not emotion, bore significant coefficients in the response model.

\begin{table}[!ht]
    \centering
    \captionsetup{width=.8\linewidth}
    \begin{tabular}{c c c c c}
        \textbf{Variable} & \textbf{Estimate} & \textbf{Standard error} & \textbf{t statistic} &  \textbf{p-value} \\
        \hline
        \hline
        \multicolumn{5}{c}{First level: RT model} \\
        \hline
        Intercept & -0.311 & 0.041 & -7.567 & \textbf{$<$0.001}\\
        Offer & -0.029 & 0.006 & -4.782 & \textbf{$<$0.001}\\
        \hline 
        \multicolumn{5}{c}{Third level: Response model} \\
        \hline 
        Intercept & 2.579 & 0.797 & 3.238 & \textbf{0.001}\\
        Offer & 3.961 & 0.524 & 7.564 & \textbf{$<$0.001}\\
        Residual RT & -2.446 & 0.326 & -7.499 & \textbf{$<$0.001}\\
        \hline
    \end{tabular}
    \caption{Estimate and significance of the coefficients in the model discussed above.}
\end{table}

\subsection*{Agglomerative Clustering of the Model Coefficients}
\begin{figure}[!ht]
    \centering
    \includegraphics[width=\textwidth]{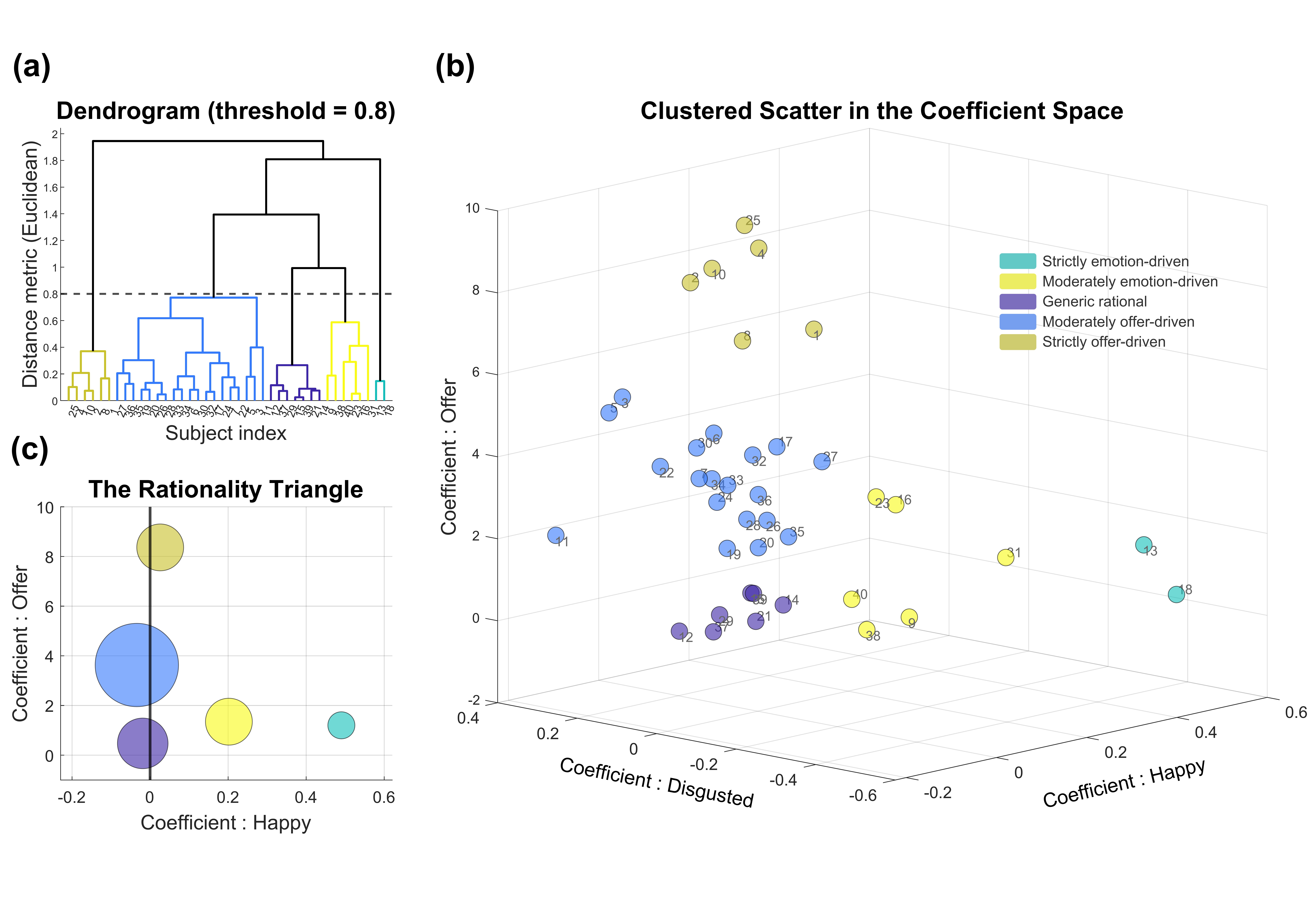}
    \caption{(a) \textbf{Identifying the groups.} The dendrogram was constructed by agglomerative clustering. Based on heuristical inspection, five primary clusters were identified (colour-coded). (b) \textbf{Clusters in the coefficient space.} Response model coefficients of two facial expressions - happy, disgusted (that is the term $(1|i\cdot e)$ in the response model) and offer (the term $(\mathcal{O}_{ibet}|i)$) in the response model are scatter-plotted. (c) \textbf{The rationality triangle.} The clusters are represented by filled circles centred at the group centroids, with radii proportional to the group sizes.}
\end{figure}
Due to the mixed-effect nature of the model, it was possible to investigate the subject-level variation of the modelling parameters. Once the coefficients of two facial expression-related variables (happy and disgusted) and normalized offer were plotted in a scatter, these formed a few obvious groups of participants which can be interpreted by the corresponding behavioural strategies (for example, prioritizing facial 
expressions strongly over financial fairness). An agglomerative clustering approach was employed after normalizing the coefficient space (using Euclidean distance metric and Ward clustering method \parencite{Murtagh2014}). 
\\
By manual verification, five broad clusters were identified from the dendrogram (Figure 4 (a)): namely, strictly driven by positive emotion displayed by the proposers (SE), moderately driven by positive emotion (ME), generic rational: participants who accepted almost all the offers (GR), strictly driven by the offer (SO) and moderately driven by the offer (MO).
\\
A comprehensive visual in Figure 4 (c) shows these five main clusters (centred at the class centroid and radius proportional to the number of members). Participants primarily focused on either of the two variables, namely, offer and emotion, during decision-making.

\subsection*{Drift Diffusion-based Insights}
\begin{figure}[!ht]
    \centering
    \includegraphics[width=.9\textwidth]{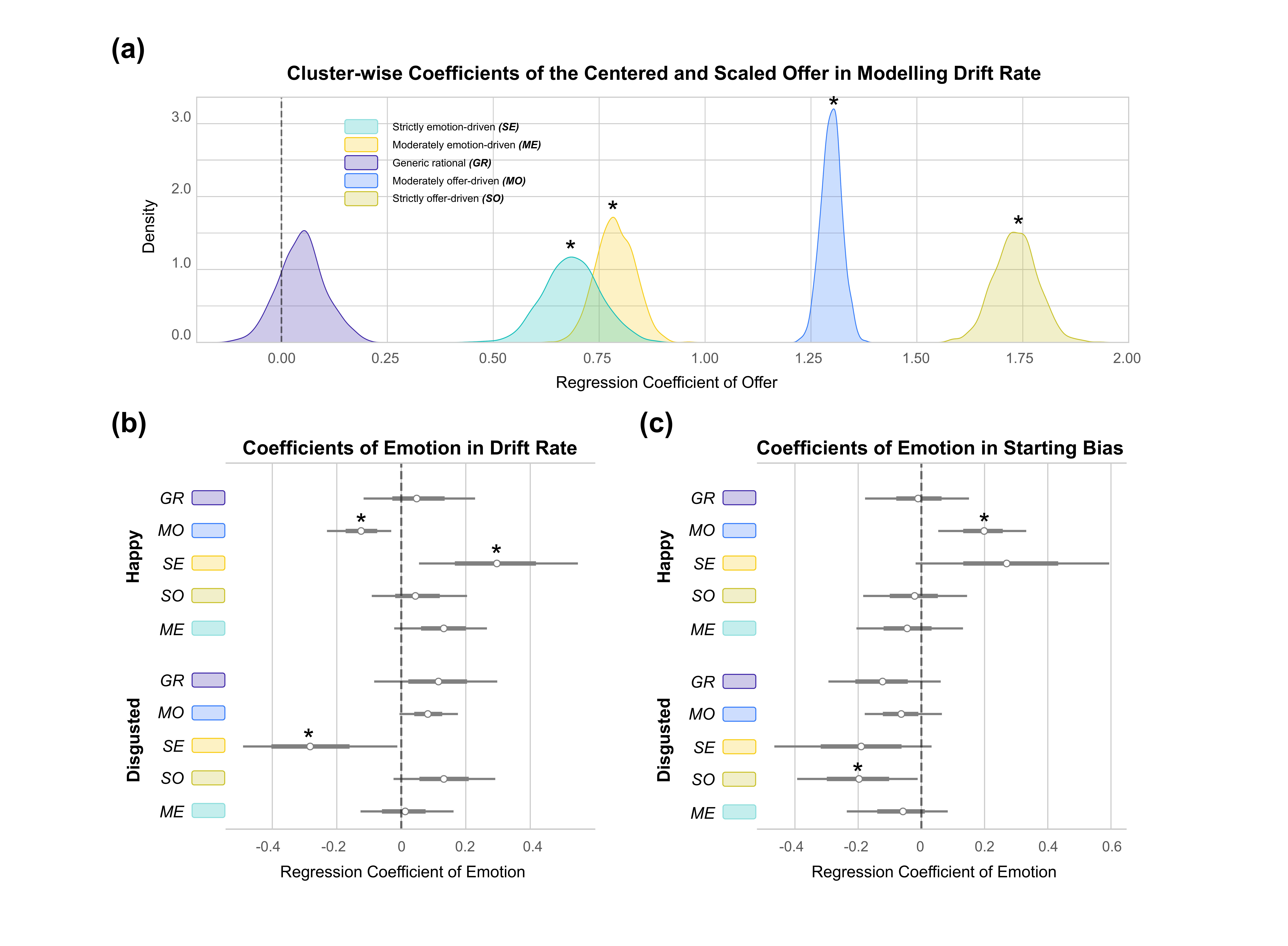}
    \caption{Drift Diffusion Model coefficients with mean and 95\% CI. \textbf{(a)} The coefficient of normalized offer in regressing the drift rate is depicted across five behavioural clusters, as identified via agglomerative clustering. 
    \textbf{(b)} 95\% CI of the intercepts corresponding to two emotions (namely, Happy and Disgusted) in modelling the drift rate are shown across the behavioural clusters.
    \textbf{(c)} 95\% CI of the intercepts corresponding to two emotions (namely, Happy and Disgusted) in modelling the starting bias are shown across the behavioural clusters.}
\end{figure}
We hypothesized that the emotion-driven clusters would be prejudiced in setting the starting point ($z$), whereas the impact of the offer's fairness would be manifested in the drift rate ($v$), which is graphically represented in Supplementary material Figure 3. The model was fitted for across all the participants with participant index and their cluster identity as random effects. In Figure 5, the distributions with their 95\% confidence intervals are reported.
\\
The coefficient of offer was found to be positive and significant for all four clusters, except for the generic rational group. Even though, it was not significantly different between group SE and ME, there was a consistent and significant increase from ME to MO and MO to SO. Notably, the offer appeared to not impact the drift rate for the GR group, consistent with the insights gained from the cognitive model.
For both emotion-driven clusters (SE and ME), the coefficients for happy and disgusted emotions in modelling the drift rate are generally positive and negative, respectively. Generally positive or negative essentially indicates that most of their distributions lie above or below zero (Figure 5 (b)). However, these effects are statistically significant only in the strictly emotion-driven group (SE).
\\
Only for SE, the starting bias was positively impacted by a happy emotion and negatively so by a disgusted emotion. However, these effects were non-significant. For MO, the coefficient of happy emotion was found to be significantly positive, whereas, for SO, it was disgusted emotion's coefficient (Figure 5 (c)). However, these emotion-based predispositions were possibly counteracted by the opposite trends in the drift rate (Figure 5 (b)).
\\
Contrary to expectation, the impact of emotion was manifested in the drift rate, and not the starting bias. The coefficient of offer in modelling drift rate was consistently positive and showed statistical significance. Its mean increased as the group placed greater emphasis on the offer (Figure 5 (a)). Potential interpretations are delineated in the Discussion section.

\subsection*{Neural Data Analysis}
Neurophysiological signal (EEG) induced by the proposer's face and their offer was collated and analysed for all participants and different clusters as constructed previously. In this section, we majorly focused on SO and E$_{\text{merged}}$ (by merging SE and ME to improve the trial size) and discussed the pertinent face-triggered and offer-triggered neural markers. Results obtained from the analysis of the other clusters can be found in Supplementary A. 

\subsubsection*{Face-epoched EEG Analysis}
\begin{figure}[!ht]
    \centering
    \includegraphics[width=\textwidth]{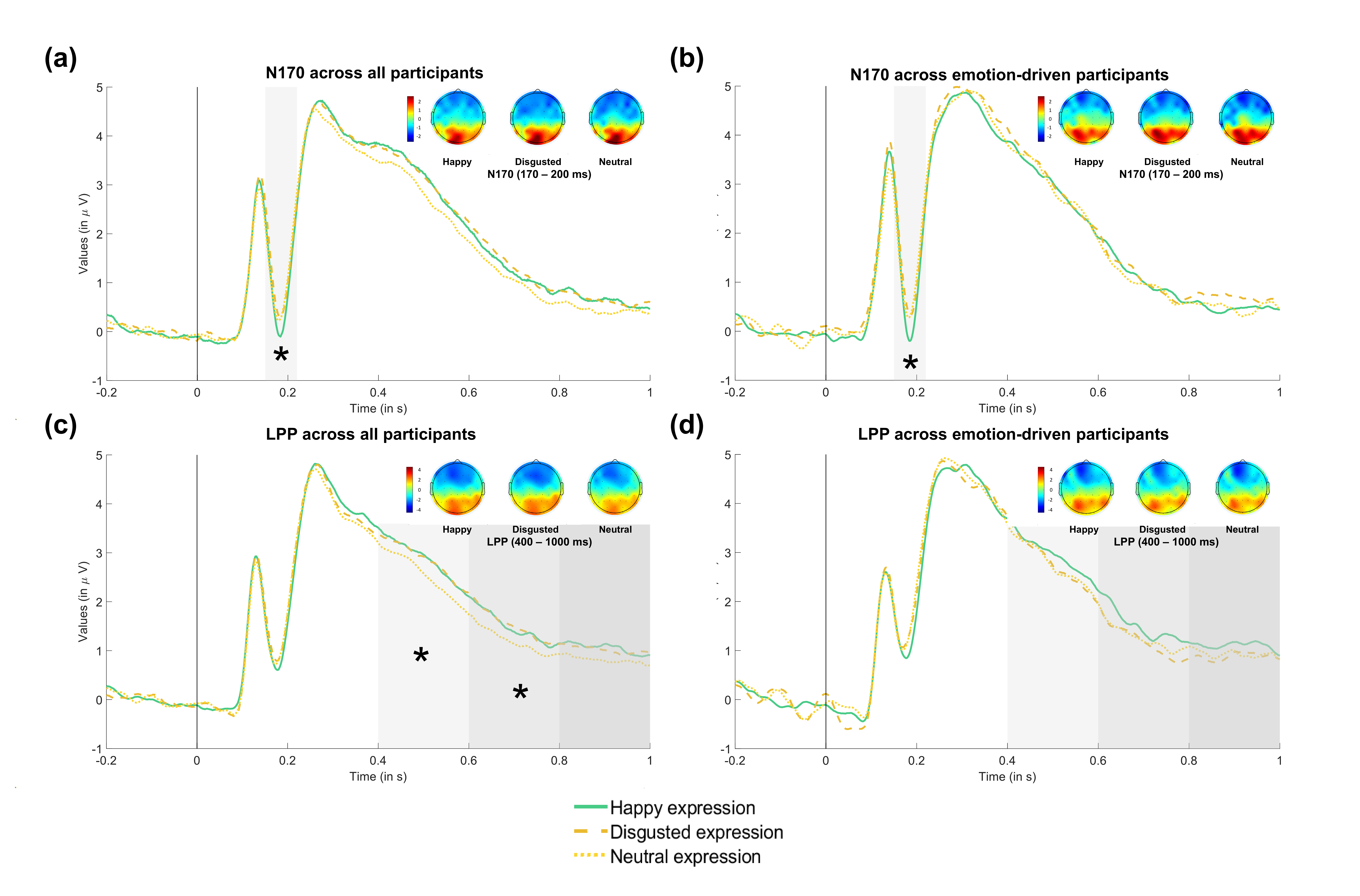}
    \caption{(a, c) \textbf{Overall ERP.} Face-triggered EEG signal for happy, disgusted, and neutral emotions across 40 participants. A significant effect of emotion on N170, early (400 - 600 ms) and mid-LPP (600 - 800 ms) was observed. (b, d) \textbf{ERP for E$_{\text{merged}}$.} For the emotion-driven participants, emotion affected only N170 and not LPP. Note that the N170 marker was investigated at parieto-occipital electrodes (PO7, PO8, P7, P8), whereas early, mid, and late-LPP were analysed at parietal and parieto-occipital electrodes (Pz, P1, P2, P3, P4, POz, PO3, PO4). (*): $p < 0.05$}
\end{figure}
We aimed to investigate the effect of differential facial expressions elicited by the proposer on the underlying cognitive processing of the participants. Using a one-way repeated measure ANOVA on the event-related potential (ERP), grouped across different electrodes for 40 (forty) subjects, we identified an effect of the proposer's emotion on the N170 peak and Late Positive Potential (LPP). 

\begin{itemize}
\item \textbf{P100}:
A prominent P100 was observed across all three emotions. However, no significant effect of the proposer's facial expression is found ($p = 0.17$). 

\item \textbf{N170}:
A pronounced negative peak was observed between 170 and 200 ms post-stimulus. The N170 peak, widely recognized as a human face-perception marker \parencite{johnston2015n170}, was affected by differential facial expressions displayed by the proposers ($F = 4.619, \eta^2_p = 0.1, p = 0.012$, Figure 6 (a)). Tukey's post hoc test suggested that the peak induced by a happy proposer was significantly more pronounced than that induced by a neutral proposer ($p = 0.003$), while its amplitude against a disgusted face ($p = 0.06$) was more similar to a neutral face. 
\\
For the emotion-driven participants (E$_{\text{merged}}$),we found a significant effect of emotion on the N170 amplitude ($F = 3.74, \eta^2_p = 0.08, p = 0.02$, Figure 6 (b)). No significant effect was observed for other clusters. 

\item \textbf{Late Positive Potential (LPP)}:
LPP is characterized by a sustained positive amplitude from 400 ms to 1000 ms post-stimulus. To gather further insight into the temporal dynamics, we analysed this time window by dividing it into three segments: early (400 to 600 ms), mid (600 to 800 ms) and late (800 to 1000 ms). 
\\
A repeated measure ANOVA on the temporal average of the ERPs in the aforementioned time windows suggested a significant effect of the differential emotions exhibited by the proposers in the early LPP ($F = 4.74, \eta^2_p = 0.1, p = 0.011$), mid-LPP ($F = 4.578, \eta^2_p = 0.1, p = 0.013$) and no significant effect in late-LPP ($F = 2.795, \eta^2_p = 0.05, p = 0.067$), as depicted in Figure 6 (c). 
The post hoc test suggested a significant difference between happy \& neutral ($p = 0.028$) and disgusted \& neutral ($p = 0.009$) in the early LPP and disgusted \& neutral ($p = 0.012$) in the mid-LPP.
\\
E$_{\text{merged}}$ did not show a significant effect of emotion on early ($p = 0.51$), mid ($p = 0.18$), or late-LPP ($p = 0.45$)(Figure 6 (d)). 

\end{itemize}

\subsubsection*{Offer-epoched EEG}
\begin{figure}[!ht]
    \centering
    \includegraphics[width=\textwidth]{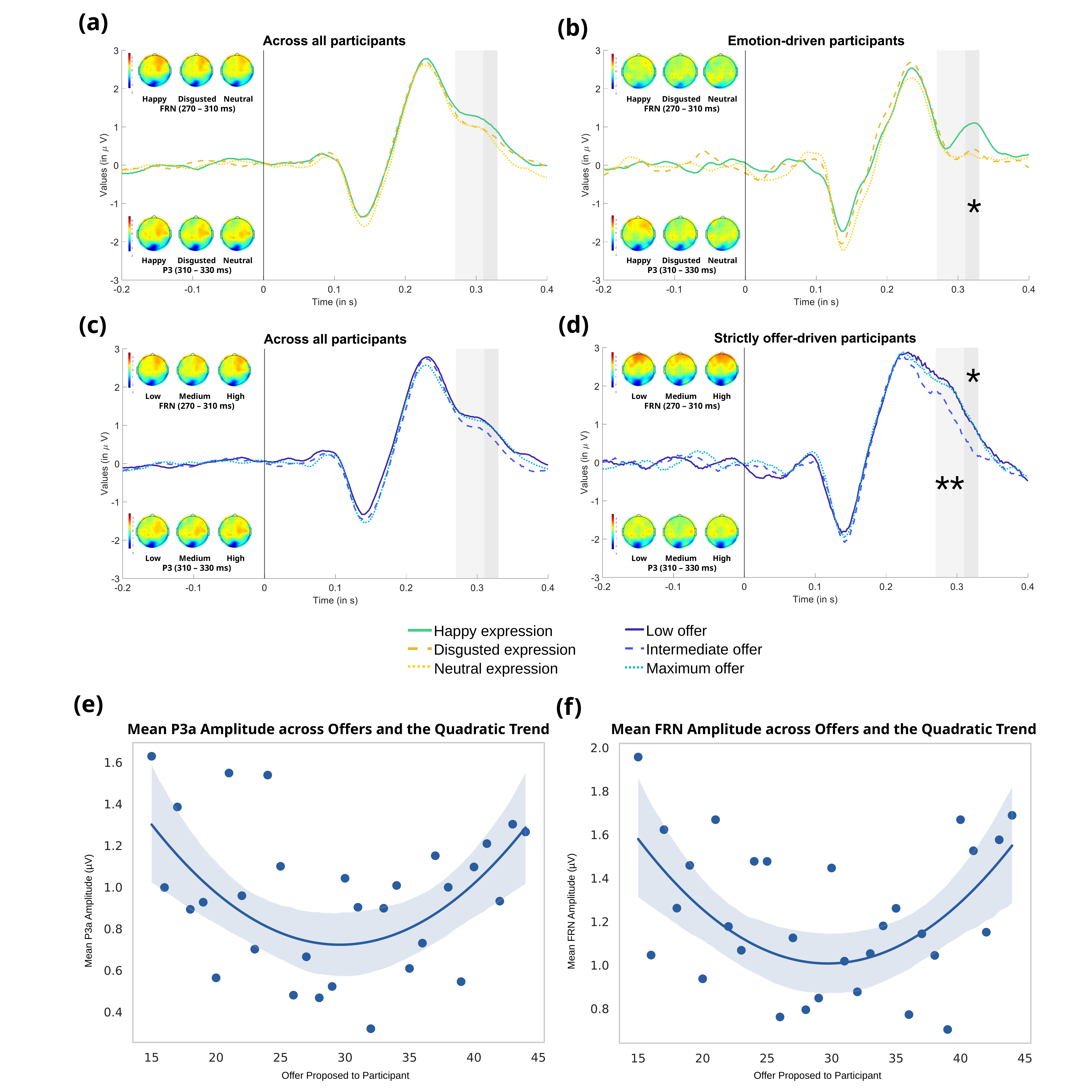}
    \caption{(a, b) \textbf{Offer-epoched ERP, grouped by emotion.} Offer-triggered EEG signal across happy, disgusted, and neutral expressions for (a) all participants and (b) E$_{\text{merged}}$. Emotion significantly modulated the P3 amplitude only for E$_{\text{merged}}$. (c, d) \textbf{Offer-epoched ERP, grouped by offer.} Offer-triggered EEG signal across low, intermediate, and maximum offer ranges for (c) all participants and (d) SO. Only for SO did the offer significantly affect the FRN as well as the P3 window. FRN (lighter shade) and P3 (darker shade) were analysed at the frontal electrodes (F1, F2, Fz, and Fcz). (*): $p < 0.05$, (**): $p <0.01$.  (e) and (f) show the \textcolor{black}{quadratic} trends in the trial-wise P3a and FRN amplitudes across offer.}
\end{figure}

The experimental design enabled us to investigate the neural processing of the participants in response to the proposer's offers, independent of their facial expressions. Trials with a reaction time lesser than 250 ms were eliminated, observed to be a biological threshold of human reaction time \parencite{jain2015comparative, Myers_2022}. 

\begin{itemize}
\item \textbf{Feedback Related Negativity (FRN)}: A negative trend was observed between 270 to 310 ms post-stimulus. Considering all the participants, no significant effect ($p=0.08$) was observed on FRN across different perceived emotions (Figure 7 (a)). No significant effect of the offer could be observed on FRN either ($p=0.1$, Figure 7 (c)). 
Comparing responses of acceptance and rejection across all the trials, we found no significant difference in FRN amplitude ($p=0.16$). 
\\
No significant effect of emotion on FRN was found for E$_{\text{merged}}$ ($p = 0.12$, Figure 7 (b)). However, a strong influence of the offer was detected for the group SO ($F =  10.74, \eta^2_p = 0.68, p = 0.003$, Figure 7 (d)). Tukey's post hoc test revealed a significant difference between low \& intermediate ($p=0.033$) and maximum \& intermediate ($p=0.047$) offers.\\
The FRN amplitudes depicted a \textcolor{black}{quadratic} trend when plotted across the offers (Figure 7 (f)), resembling the previously reported U-shaped effect \parencite{Rodrigues2024}.

\item \textbf{P3}: For both P3a and P3b, no prominent peak was evident in the ERP. Therefore, P3a was defined to be a sustained positive trend between 310 and 330 ms. No significant effect of emotion ($p=0.056$) or offer ($p=0.14$) was found on P3a across all subjects (Figure 7 (a) \& (c)).
\\
For the group E$_{\text{merged}}$, the effect of emotion on the P3a amplitude turned out to be significant ($F = 4.8856, \eta^2_p = 0.41, p = 0.024$, Figure 7 (b)), with a happy face invoking the highest amplitude. A post hoc test revealed a significant difference ($p=0.049$) between happy and disgusted emotions. SO showed a significant effect of offer on P3a ($F =  4.612, \eta^2_p = 0.48, p=0.038$, Figure 7 (d)). Tukey's post hoc test revealed no extra level of significance. The P3b was not well-defined overall or in any of the behavioural clusters. We observed an increasing positivity in between 330 and 400 ms (Supplementary material Figure 7).\\
Similar to the FRN amplitude, the P3a amplitudes depicted a \textcolor{black}{quadratic} trend across the offers proposed, with medium offers (those near 30) invoking the least P3a response. This observation resembles the recent findings by Rodrigues and colleagues for the P3b component \parencite{Rodrigues2024}.
\end{itemize}


\section*{Discussion}
The primary objective of the study was to analyse the effect of facial expression, connoting two specific fundamental emotions of opposing valence (``happiness" and ``disgust") on rational inference in a time-bound condition, with the preliminary hypothesis that the decisions would primarily be driven by the economic fairness, with a partial/small effect of the perceived emotion \parencite{ferra2021}. 
The results validate the expectation that offer fairness is the strongest factor determining response as well as response promptness, which has been observed in multiple cases. However, we obtain significant effect of emotion on the response promptness or the reaction time, but not on the acceptance rate.

\subsection*{Fairness as the Dominant Variable}
\subsubsection*{The Acceptance Rate}
The overall results, using ANOVA analysis and the generalized linear modelling approach, seem to suggest that economic fairness was the strongest candidate in shaping the responder's feedback (Figure 2 (d)). This result reinforces the prior findings \parencite{ferra2021, mussel2013value}, claiming that the fairness of the offer firmly steers the decision. These findings, however, contradict earlier studies showing a substantial main effect of the facial expression \parencite{mussel2013value, mussel2014smiling, weib2021}. Two possible factors might be contributing to this dichotomy.
\\
Firstly, the participant pool in most of these studies has been quite heterogeneous insofar as the age range and educational level are considered \parencite{mussel2013value}. On the other hand, the current study focused on a particular domain of the spectra - freshman college students 18-20 years of age. The effect of education - particularly numeracy training - is often correlated with rationality and framing avoidance \parencite{Fan2017_edu}.
\\ 
Secondly, the offer size seems to play an essential role in setting the economic incentive for the participants. The study by Mussel and colleagues utilised a small-scale monetary offer (distribution of 12 cents), which contrasts with the scale used by Ferracci and colleagues (a total of 100 cents), which is similar to our study (\rupee100). A high-magnitude offer may induce more robust attention towards gauging monetary fairness than the emotional features or smaller economic incentives may enhance the credibility of the narrative \parencite{ferra2021}.
\\
Additionally, we tested whether the attractiveness of the proposers' faces influences the reaction time and the acceptance rate. The mixed-effecting modelling indicates that the coefficients of the attractiveness rating are not significantly different from zero (Supplementary Table 11), thus refuting the possibility of a ``handsome bias" occurring in the current context \parencite{ma2015}.

\subsubsection*{The Reaction Time}
Even though it is challenging to infer the cognitive implication of a delayed reaction fully, the latency speaks for the processing speed, which, in turn, can be attributed to the strength or quality of the information in the stimuli, cognitive load, encoding or retrieval delay. In tasks designed for identifying emotional features in facial expressions, participants spend more time on the emotional faces than the neutral ones, possibly due to an excess time required to process the emotional stimuli on top of the face stimuli \parencite{emotprocess2012}.
\\
The current study fosters this theory, as a higher reaction time is observed while playing against a proposer displaying emotional expression (Figure 4 (e)). Moreover, we find that the latency is the highest against a disgusted expression. Also, the difference between the latencies against a happy and a disgusted proposer is inflated for the unfair range of offers. 

Another prominent result indicates that acceptance is less time-consuming than rejection (Figure 2 (c)), an observation we term the ``Speed-Acceptance tradeoff", as it resembles the classical speed-accuracy tradeoff in cognitive science \parencite{Kar_lar_2014}. This can be explained by the assumption that rejection is a result of inhibiting impulses related to self-interest, and thus, prolonged cognitive deliberation \parencite{Mussel2013cogneed}.

\subsubsection*{Neural Markers-based Interpretations}
Most earlier studies exploring neural correlates of emotion in connection to ultimatum games have broadly focused on the offer-evoked neurophysiological features \parencite{mussel2014smiling, mussel2022neural}. 
In contrast, the experimental design enabled us to investigate the neural markers triggered by the face and the offer stimuli separately. In the first segments of the trials, when the participant merely observes the proposer's face for a second, we show a significant effect of the emotional expression on the N170 and LPP components (early and mid-LPP). 
\\
The N170 marker indicates an early subconscious attention allocation towards facial features \parencite{schindler2020attention}. A smiling face inducing a strong N170 (Figure 6 (a)) possibly supports the ``happy face superiority" hypothesis \parencite{happysuperior2013}. However, the consensus amongst researchers denotes that the N170 can be elicited in covert and overtly attended stimuli. 
\\
The LPP (400 to 1000 ms post-stimulus) denotes long-term attention allocation and deliberate processing. Both emotions evoke a higher amplitude of early and mid-LPP (400 - 800 ms) than the neutral expression (Figure 6 (c)), suggesting prioritization of stimuli with higher emotional valence and possible induction of cognitive load later impacting the reaction time, consistent with standard literature \parencite{schindler2020attention}. 
\\
These results indicate that the participants register facial expressions before responding to the offer. Therefore, while interpreting the overall null effect of emotional expression on the acceptance rate, we must acknowledge the long-term attentiveness towards the emotional components in the stimuli - which is also reflected via the reaction time analysis. However, during the final decision, the participants deliberately focused only on economic fairness, overriding the effect of facial expressions developed prior to that. The effect of emotion being lost in the late-LPP zone (Figure 6 (c)) is a probable index of the reduced prioritization of facial expressions.
\\
While analysing the offer-triggered ERP, we focused on the FRN marker, which is credited to being associated with negative social feedback, punishment tendencies and cognitive conflict \parencite{polezzi2008mentalizing, mussel2014smiling, weiss2020smiling, zhong2019midval} and  P3a (or, sustained positivity in the absence of a pronounced peak), frequently associated with stimulus-evaluation and salience \parencite{polich2007updating, P3_Peterburs2017, P3_Mansor2021}.
\\
The ERP, grouped by the emotional expressions the participant perceived before receiving the offer, shows a less pronounced FRN amplitude against a happy proposer than both disgusted and neutral ones (Figure 7 (a)). Offer range-grouped ERP reveals that the FRN activity is comparatively poorer when the participant receives a fair offer (in the range 35 - 44) than when they receive an offer in the range 15 - 34 (Figure 7 (c)). Surprisingly enough, we fail to observe a significant effect of either emotional expression or offer magnitude on the FRN amplitude, even though the mean ERP trends are consistent with the literature. However, the FRN amplitude showed the \textcolor{black}{previously found quadratic} trend with the mid-value offers eliciting the highest negativity. This has previously been explained as a result of heightened cognitive conflict between the self-interest component and the emotive process (discussed previously). \textcolor{black}{Though \parencite{mussel2014smiling} have reported a cubic trend of FRN activity against offer, by systematically modelling this, we found quadratic trend to be better fitted as shown in Supplementary material Table 11. One possible explanation of this could be due to the fact that, in the paradigm of \parencite{mussel2014smiling}, the receiver could get 66\% of the offer where in our case it was 44 \% so we are observing only a subpart of the whole cubic curve which is quadratic.}
\\
Not finding a significant effect of offer magnitude on the FRN amplitude possibly hints at the necessity to explore and deliberate on different ``punishment" markers or components related to inhibitory control - not only in the time domain but also in the frequency domain, e.g., the mid frontal theta \parencite{Nayak_2019, Rodrigues2024}. Plausible reasons behind not observing a significant effect on FRN include - (i) inadequacy of trials ending in rejection, as FRN is often associated with negative external feedback which, in the current context, will be rejected \parencite{wang2020_frn}, (ii) involvement of hypothetical monetary stakes \parencite{ferra2021}, (iii) involvement of a `norm abidance' tendency over `unfairness aversion' \parencite{vavra2018expectUG}.
\\
Overall, a distinguishable P3 signal (neither P3a nor P3b) was not detected in this study. Considering that not many of the earlier prominent studies have explored the P3 component \parencite{mussel2014smiling, weiss2020smiling} and earlier failures to observe response-based amplitude modulation \parencite{P3_Peterburs2017}, a possible explanation may be attributed to individual differences in stimulus evaluation, widely recognized to modulate the P3 amplitude. Notably, we could only observe an increasing positivity between 330 and 400 ms (Supplementary material Figure 7). This trend possibly would lead to a P3b component at a latency greater than 400 ms. Given the incomplete P3b, we abstained from further statistical analyses which could lead to fallacious insights. As will be discussed in the subsequent section, we observe noteworthy trends in the P3a signal within different behavioural clusters, categorised by either emotion or offer-based grouping. Additionally, in rapid response paradigms like the current one, observing a P3b can be difficult for its longer latency (300 - 600 ms) and more careful stimulus evaluation. On the contrary, the P3a indicates a covert response via orienting the attention and has a shorter latency \parencite{polich2007updating}. Interestingly, we could replicate the \textcolor{black}{quadratic} trend of the P3a amplitude, in resemblance with the findings on P3b \parencite{Rodrigues2024}. This finding possibly suggests that the observed effect may not be exclusive to one subcomponent, but rather reflects a broader modulation of the P3 complex as a whole.

\subsection*{A Cognitive Insight into the Behavioural Clusters}
We grouped the participants broadly into five behavioural clusters (SE, ME, GR, MO, SO) and explored their responses and strategies. The multilevel mixed-effect linear model not only reinforces the repeated measure ANOVA inferences but also unveils a novel avenue to approach this paradigm. Surpassing an overt generalization, this work dives into individual strategies and possible interpretations. 
\\ 
Prior to discussing the underlying decision-making mechanism for each of these individual groups, it is imperative to characterise the drift rate and its cognitive interpretation.
The DDM is an accumulator model that samples from a noisy source of information until it reaches a predefined decision threshold \parencite{ratcliff2008ddm}. Thus, a natural question arises: what information (and where from) is the DDM ``accumulating" when applied in the current context?
While playing the ultimatum game, the participants have two parallel streams of strategy or intention \parencite{wei2022dual}. The first intent is conceptualized through a system engaged in maximizing profit and self-gain, which should always push the decision-making towards accepting the offer. Simultaneously, the participants consider the fairness of the offer, whether the agent is trustworthy, and so forth. The drift rate functions as an index depicting the priority of either of the strategies, given a particular set of conditions (emotion and offer). The coefficients of these conditions while modelling the drift rate indicate their linear relationship with the drift rate, which, in turn, depicts whether profit is prioritized over fairness considerations. 
Interpreting the starting bias is relatively straight-forward. It simply indicates whether the emotion perceived prior to encountering the offer sets a predisposition towards acceptance or rejection.

\subsubsection*{Participants Driven by Positive Emotion}
SE and ME respond positively to a happy expression and unfavourably to a disgusted one. They prefer not to focus too strongly on economic fairness,as indicated by the DDM analysis (the coefficients of offer in modelling the drift rate for these groups are considerably lower than the other groups (Figure 5 (a)). Furthermore, the drift-diffusion modelling suggests that the emphasis on the positive emotion does not stem from an initial bias prior to the participants observing the offer; Rather, this is manifested in the drift rate. At a constant offer, a lower drift rate against a disgusted expression (Figure 5 (b)) implies a heightened prioritization of the second system, scrutinizing fairness consideration and emotional experience \parencite{Sanfey2008dual}. This manifestation is stronger in SE than ME, as expected.
\\
The neural markers were investigated after merging these two clusters (thus, forming E$_\text{merged}$) to improve the trial size. The face-epoched ERP shows a significant effect of emotion on N170 (highest amplitude against a happy expression, followed by disgusted and neutral), the strongest amongst all the clusters. However, we fail to observe any modulation of the late positivity. Offer-epoched ERP, grouped by offer, reveals no influence on either FRN or P3. When grouped by emotion, however, we observe a reduced FRN (though not significant) and a significantly strong P3a amplitude against a happy expression. 
\\
In conjunction with the DDM outcomes, we infer that the participants prioritizing the positive emotion are fundamentally sensitive to a happy face (as indicated by a strongly modulated N170). Possibly, their long-term attention is uniformly allocated for each of the expressions they perceive (thus, an undifferentiated LPP and hardly an emotion-induced initial bias in the drift-diffusion model), and the disgusted expression is discriminated only after they initiate processing the offer. At this stage, the offers from a happy proposer are deemed less inequitable (reduced FRN amplitude \parencite{wang2020_frn}). More importantly, attention allocation is strongly impacted by a happy expression, in accordance with the ``happy face superiority effect" \parencite{happysuperior2013, Stuit2025}. This reflects in the drift rate coefficients corresponding to emotion.

\subsubsection*{Participants Driven by Economic Fairness}
SO and MO prioritize the fairness of the proposed monetary split. Thus, the probability of acceptance increases with a higher offer. As evident in Figure 4 (c), these groups maintain a neutral stance with respect to the perceived facial expression. A strongly positive coefficient of offer (consistent increase from ME to MO and SO) in modelling drift rate directly depicts how increasing fairness shifts the priority of the latent decision variable on maximizing profit as it elevates the drift rate (Figure 5 (a)). 

Interestingly, a subtle effect of emotion can be observed in the corresponding cluster-specific intercepts for these two groups. A happy face develops a starting bias towards acceptance for MO, while, for SO, a disgusted face sets a predisposition towards rejection. This bias, however, is nullified by the opposite impact of emotion in modelling the drift rate (Figure 5 (b)), thus, depicting no overall effect of emotion. Explaining this mechanism is not straightforward and demands further analysis.

In the face-epoched ERP, no significant modulation was found for either MO or SO. Interestingly, we did not observe a pronounced FRN for these two groups either. By using the same FRN time window as for the other clusters, a significant effect of the offer's fairness was found. Similarly, in the absence of a clear P3 peak, the sustained positivity between 310 ms and 330 ms post-stimulus showed a significant effect of the offer. Note that the amplitude of the signals corresponding to the most and the least fair offer ranges is similar and is higher than that corresponding to the ambiguous offer range (in other words, the intermediate range), as indicated earlier through the \textcolor{black}{quadratic} trend in the FRN amplitude (Figure 7 (e)). 
Previous studies have shown that both unfair and fair offers elicit similarly high levels of arousal and salience, greater than those evoked by ``sub-fair" or mid-range offers \parencite{Hu2014offersalience}. Also, such mid-value offers are exhibited to elicit greater cognitive conflict. Both these attributes can be explained by less pronounced positivity over the P3a window and greater negativity in FRN, respectively.

\subsection*{Limitation}
We also have some limitations of our study. Firstly, in the current study, a jitter between the face stimulus and the offer presentation screen, primarily to constrain the total time of the experiment. Though this setting allows for a more naturalistic framework that resembles real-life bargaining scenario, it may lead to unwanted variability in the offer-induced electrophysiological signal. Although, we have analysed Contingent Negative Variation (CNV) as a marker of anticipation and could not find any trends suggesting negligible anticipatory effects (For details see Supplementary material G).Secondly, the paradigm involved hypothetical economic incentives, as reported by Ferracci and colleagues \parencite{ferra2021}.  Smaller, and perhaps more realistic, incentive could invoke a greater degree of reliability and engagement with the task amongst the participants, as was the case for Mussel et al. \parencite{mussel2013value}.
Thirdly, the absence of human agents might have impacted the acceptance behaviour, as unfair offers from a regular computer are less often rejected than those from a human agent \parencite{agent_comp_2019}.
Fourthly, in contrast to most of the earlier studies, we failed to find a significant FRN modulation, though we found neurophysiological evidence indicating that the participants registered the emotion. 
Additionally, the P3b component (expected to be found between 400 and 500 ms) could not be analysed as the participants had mostly responded before 500 ms, and the EEG signal would be hard to interpret due to motor activity-related noise.
Another limitation of the current study is the sample size and some of the clusters being relatively small. Though the proposed strategy landscape has stayed consistent under multiple modelling frameworks (refer to Supplementary Material B), pre-registered studies (similar to \parencite{rodrigues2022second,Rodrigues2024}) should be conducted in the future to confirm these findings with a larger sample size. 

\section*{Summary}
To summarize, this study emphasized the stronger influence of economic fairness over overt facial expressions in a social decision-making task using neural correlates and computational modelling.
These analyses could recognize the multitude of strategies that the participants explored and provided a finer insight into the cognitive mechanism underlying each of these strategies through the lens of drift diffusion-based modelling and strategy-specific ERP studies.
While interpreting and furthering the findings, it is imperative to note that the study involved only heterosexual male first-year undergraduate pan-Indian students in the age range 18 - 20. These parameters were treated as invariants as the sex and age of the agents were found to impact decision outcome \parencite{bailey2013age} \parencite{ma2015} \parencite{ferra2021}.
Future directions include registering the clustering-based approach on a larger sample size across age, gender, and demographics to check the consistency of the proposed ``rationality triangle". Various decision-making models and neuroimaging data can be interpreted via this approach. Furthermore, the impact of a variable economic incentive is worth investigating. 

\section*{Data Availability Statement}
The codes and the processed data are made available on GitHub at
\url{https://github.com/MR-dot-15/UG_data}. All the data collected will be made available on request.

\section*{Competing Interests}
The authors declare no competing interests.

\section*{Author Contribution}
\textbf{R.M}: Conceptualization, Formal Analysis, Methodology, Software, Validation, Writing - Original draft preparation. \textbf{S.C}: Conceptualization, Formal Analysis, Methodology, Software, Validation, Writing - Original draft preparation. \textbf{K.D}: Conceptualization, Methodology, Funding Acquisition, Supervision, Validation, Writing - Original draft preparation.   

\section*{Acknowledgements}
The authors are grateful to Dr. Satyaki Mazumdar for his enormous support and insights during the project and for helping us with the DDM. We will also acknowledge Subhajit Das for helping us with some parts of the data collection.  We also thank anonymous reviewers for their suggestions to improve this manuscript.

\printbibliography

\end{document}